\documentstyle[12pt]{article}
\def\be{\begin{equation}}
\def\ee{\end{equation}}

\begin{document}

\title{Dynamical systems where time is a quantum group
and quantum ergodicity}

\author{S.V.Kozyrev}

\maketitle

\begin{abstract}
We define dynamical systems where time is a quantum group.
We give the definition of quantum ergodicity for the introduced
dynamical system with noncommutative (or quantum) time,
and discuss the examples.
\end{abstract}

\section{Introduction}

In the present short note we consider dynamical systems where time
is a quantum group and introduce the definition of quantum ergodicity.
The paper is based on the papers \cite{Kozy}, \cite{KozReg}.

Quantum groups, see \cite{Skl}--\cite{KozAAV},
is a celebrated example of noncommutative geometry.

The main observation, used in the present paper, which corresponds to
noncommutative ergodicity, is the following.
We consider tensor products of representations of noncommutative algebra
of functions on the quantum group
(tensor products of representations of the algebra of noncommutative matrix
elements of $SU_q(2)$ were considered in
\cite{Kozy}, \cite{KozReg}, \cite{LesR}) and observe, that the products
of representations stabilize: if we take some (finite) number
of nontrivial representations in the product, then the tensor products
of the obtained representation $\Pi$ and of an arbitrary unitary representation
will be unitarily equvalent to $\Pi$.

We also discuss the analogue of this property which corresponds to
such important subject of mathematical physics as quantum Bethe anzats
in the noncommutative inverse problem method \cite{FTS}, \cite{FaT}.

The relation of the discussed above property and ergodicity is the following.
If we consider the dynamical system of shifts on a group,
then the shift by $g_0$ of the measure $\mu$ will correspond to
the tensor product of GNS representations of algebra of functions on the
group, corresponding to
the measure $\mu$ and to $\delta$--like measure at $g_0$.
Then the existence and uniqueness of the representation,
invariant with respect to shifts, corresponds to ergodicity.

\section{Dynamical system where time is a quantum group}

In the present section we discuss the property of ergodicity
and introduce the noncommutative analogue of ergodicity.

Abstract dynamical system is the triple $(M,\mu,\phi)$,
where $M$ is the set with measure $\mu$,
and the map $\phi:M\to M$ conserves the measure $\mu$.

Abstract dynamical system $(M,\mu,\phi)$
is ergodic, if for integrable function $f$
the time average is equal to the space average:
$$
\lim_{N\to\infty}{1\over N}\sum_{i=0}^{N-1}f(\phi^i x)=\int f(x)d\mu(x)
$$

Abstract dynamical system is ergodic if and only if the arbitrary
measureable invariant set $S$ has the measure $\mu(S)=0$ or
$\mu(M\backslash S)=0$.

The well known example of ergodic dynamical system is the shift
of the torus with irrational angle.

In the present note we construct analogous example of
ergodic action of quantum group.

We discussed above dynamical systems with classical time
(time is the semigroup of natural numbers). Dynamical systems with non
classical time, when time is an arbitrary (semi) group, were considered.

Discuss now the definition of dynamical system, where time
lies in a quantum (semi) group $G$. Denote $Fun(G)$ the Hopf $*$--algebra
of polynomial functions on quantum group $G$
(for quantum semigroup $Fun(G)$ is a bialgebra).
Consider unitary representation $T$ of $Fun(G)$ in the Hilbert space
${\cal H}_{T}$.

Consider the family ${\cal A}$ of unitary representations
of $Fun(G)$ containing representation $T$.
In principle ${\cal A}$ may contain not all the
representations of the described above type.

The important example of the representation of $Fun(G)$
is the GNS representation generated by some state $\phi$
on $Fun(G)$.

There is an operation of tensor product $TS$
of representations $T$ and $S$ in ${\cal A}$:
$$
TS (a)= T\otimes S(\Delta a),\qquad {\cal H}_{TS} =
{\cal H}_{T}\otimes {\cal H}_{S}
$$
Here $\Delta$ is the coproduct operation in the mentioned above
Hopf algebra (or bialgebra).

\bigskip

\noindent{\bf Definition}\qquad
{\sl We call the representation $T$ of $Fun(G)$
invariant with respect to the representation $S$, if $ST$ is unitarily
equivalent to $T$.}

\bigskip

If $T$ is the GNS representation of functions over (non--quantum) group $G$,
generated by measure $\mu$,
and $S$ is $\delta$--like representation of functions over group,
correspondent to $g_0\in G$,
then the definition above reduces to the definition of
measure $\mu$ on the group, invariant with respect to shift by $g_0$.

\bigskip

\noindent{\bf Definition}\qquad
{\sl We call the pair $(S,T)$ of unitary representations
of algebra of functions on quantum group
the noncommutative dynamical system of shifts on
the quantum group, if the tensor product $ST$
is unitarily equivalent to $T$.}

\bigskip

We see that in the definition above there is no significant difference between
the noncommutative transformation $S$ and noncommutative measure $T$.

We also will say that if
$$
ST=T\qquad \forall S
$$
then $T$ is left invariant representation (the analogous definition
$TS=T$ of right invariant representation is obvious).

Now, using the definitions above, we are ready to formulate the
definition of noncommutative ergodicity.

\bigskip

\noindent{\bf Definition}\qquad
{\sl We say that the action of representation $S$ by left shifts
on the quantum group is ergodic, if there exists the left (and right)
invariant representation $T$ with respect to shifts by all representations
of $Fun(G)$, and this representation is unique up to unitary equivalence.}

\bigskip

We also will say, that the representation $T$ is ergodic with respect to
the family ${\cal A}$ of unitary representations, if it is invariant
with respect to shifts by representations from ${\cal A}$,
and $T$ is unique up to unitary equivalence.

\section{Examples of ergodic actions of quantum groups}

In the present section we discuss some examples of dynamical systems
with noncommutative time, and some examples of quantum ergodic systems.

It is easy to see that every quantum group defines dynamical system
with noncommutative time and therefore the definitions of the previous
section are nontrivial.

Let us discuss a nontrivial example of ergodic dynamical system
with noncommutative time.

In \cite{Kozy}, \cite{KozReg} it was proven that, if we have three nontrivial
unitary representations $A$, $B$, $C$ of the Hopf algebra of matrix elements
of quantum group $SU_q(2)$ (nontrivial means that the algebra of functions
in the representation is noncommutative), then the tensor product
$$
ABC=\Pi
$$
does not depend on $A$, $B$, $C$ (the class of unitary equivalence
does not depend), and for arbitrary unitary representation $S$ one has
\be\label{ergodicity}
S\Pi=\Pi S=\Pi
\ee
Using the discussion of the present note, we see that
the result (\ref{ergodicity}) means that the action of
quantum group $SU_q(2)$
by left and right shifts is ergodic, and the invariant representation
(the noncommutative analogue of the invariant measure) is the
representation $\Pi$.

Another interesting example of dynamical system with noncommutative time
is given by the quantum Bethe ansatz, used in the quantum inverse
problem method \cite{FTS}, \cite{FaT}. For the quantum Bethe ansatz
the role of noncommutative time is played by the Yangian bialgebra, and
the invariant representation is constructed by the quantum Bethe ansatz.
It would be interesting to investigate the quantum ergodicity
of the representation corresponding to the quantum Bethe ansatz.

Another one interesting example of noncommutative ergodic dynamical
system is given by the central limit theorem of probability theory.
Actually in this case we have to use the construction which is a
deformation of the considered in the previous section.

The noncommutative algebra in this case is given by
the sequence of Heisenberg algebras ${\cal A}_n$ with $n$ of degrees of
freedom, generated by selfadjoint elements $Q_i$, $P_j$ with the relation
$$
[Q_i,P_j] = i\hbar \delta_{ij}
$$
where the Planck constant $\hbar$ is a real number.

Instead of the coproduct we use the following sequence of maps
$$
\Delta_{n}:{\cal A}_1 \to {\cal A}_n
$$
$$
\Delta_{n}:Q\mapsto {1\over\sqrt{n}}\sum_{i=1}^{n}Q_i,\qquad
P\mapsto {1\over\sqrt{n}}\sum_{i=1}^{n}P_i
$$
The tensor product of states $\phi_i$ over ${\cal A}_1$ will be given by
$$
\prod_{i=1}^{n}\phi_i (X)=\otimes_{i=1}^{n}\phi_i (\Delta_n X)
$$

The central limit theorem in this language takes the following form:
the limit of the tensor products $\lim_{n\to \infty}\prod_{i=1}^{n}\phi_i$
of the states with mean zero $\phi_i(Q)=\phi_i(P)=0$ and unit covariation
is equal to the Fock, or vacuum for the harmonic oscillator,
state for ${\cal A}_1$.

\bigskip

\centerline{\bf Acknowledgements}

\smallskip

The author would like to thank I.V. Volovich for discussions.
This work has been partly supported by INTAS YSF 2002--160 F2,
CRDF (grant UM1--2421--KV--02), and The Russian Foundation for
Basic Research (projects 02--01--01084 and 00--15--97392).


\begin{thebibliography}{99}
\bibitem{Kozy} S.V.Kozyrev, Teor.Math.Phys., 101(1994)No2
\bibitem{KozReg} S.V.Kozyrev, Doklady RAS, 343(1995)N4

\bibitem{Skl} E.K.Sklyanin,
About some algebraic structures related to the Young--Baxter equation,
Functional analysis and applications, 16(1982)N4.
\bibitem{Wor} S.L.Woronowich,   Comm.Math.Phys. 111(1987)N4,p.613,
Publ.RIMS Kyoto Univ. 23(1986)p.117
\bibitem{FRT} L.D.Faddeev, N.Yu.Reshetikhin, L.A.Takhdadjan,
Algebra and analysis 1(1988)p.129
\bibitem{ArVqGPnAG}  I.Ya.Arefeva, I.V.Volovich,
Quantum Group Particle and Non-Archimedean Geometry,
Phys.Lett.B, 268(1991)179
\bibitem{Mukhin1} E.E.Mukhin, The Young--Baxter operators and
noncommutative de Rham complexes, Math. Izvestia,
58(1994)N2, pp.108--131.
\bibitem{KozAAV}  L.Accardi,  I.Ya.Arefeva,  S.V.Kozyrev,  I.V.Volovich,
The master field for large N matrix models and quantum groups.
Modern Phys.Lett.A, 11(1995)p.2341-2345

\bibitem{LesR}  A.Lesniewski,  M.Rinaldi,
Tensor products of representations of $C(SU_{q}(2))$
J.Math.Phys. 34(1993)N1,p.305

\bibitem{FTS} L.D.Faddeev,  L.A.Takhdadjan, E.K.Sklyanin,
Quantum inverse problem method I. Theor.Math.Phys. 40(1979)N2
\bibitem{FaT} L.D.Faddeev,  L.A.Takhdadjan,
Quantum inverse problem method and XYZ--model of Heisenberg,
Math. Uspekhi 34(1979)N5

\end{thebibliography}
\end{document}